\documentclass[pra,twocolumn,epsfig,rotate,superscriptaddress,showpacs]{revtex4}
\usepackage{graphicx}
\usepackage{epstopdf}
\usepackage{amsfonts}
\usepackage{amssymb}
\usepackage{amsmath}
\usepackage{subfigure}
\usepackage{prettyref}
\usepackage{float}
\usepackage{amsmath}
\usepackage{amssymb}
\usepackage{esint}

\begin{document}
\title
{Hierarchical Theory of Quantum Adiabatic Evolution}

\author{Qi Zhang}
\affiliation{College of Science, Zhejiang University of Technology,  Hangzhou 310014, China}
\affiliation{International Joint Research Laboratory for Quantum Functional Materials of Henan Province and School of Physics and Engineering Zhengzhou University, Henan 450001,China}

\author{Jiangbin Gong}
\affiliation{Department of Physics and Centre for Computational
Science and Engineering, National University of Singapore, 117542,
Singapore} \affiliation{NUS Graduate School for Integrative Sciences
and Engineering, Singapore 117597, Singapore}

\author{Biao Wu}
\email{wubiao@pku.edu.cn}
\affiliation{International Center for Quantum Materials, School of Physics, Peking University, Beijing 100871, China}
\affiliation{Collaborative Innovation Center of Quantum Matter, Beijing 100871, China}

\date{\today}
\begin{abstract}
Quantum adiabatic evolution  is a dynamical evolution of a quantum system
under slow external driving. According to the quantum adiabatic theorem,
no transitions occur between non-degenerate instantaneous eigen-energy
levels in such a dynamical evolution. However, this is true only when the driving rate is infinitesimally
small. For a small nonzero driving rate,  there are generally small transition
probabilities between the energy levels. We develop a classical mechanics framework to address
the small deviations from the quantum adiabatic theorem order by order.
A hierarchy of Hamiltonians are constructed iteratively with
the zeroth-order Hamiltonian being determined by the original system Hamiltonian.
The $k$th-order deviations are governed by a $k$th-order Hamiltonian,
which depends on the time derivatives of the adiabatic parameters up to
the $k$th-order. Two simple examples,  the Landau-Zener model and a
spin-1/2 particle in a rotating magnetic field, are used to illustrate
our hierarchical theory.   Our analysis also exposes a deep, previously unknown connection between classical adiabatic theory and quantum adiabatic theory.
\end{abstract}

\maketitle
\section{Introduction}

Quantum evolution under external adiabatic driving has been of fundamental interests to physicists.
Born and Fock proved the quantum adiabatic theorem shortly after the discovery of the
Schr\"odinger equation~\cite{adiabatic}.   This theorem states that
no transition occurs between instantaneous eigen-energy levels in a system under adiabatic driving.
However, this is only true when the external driving is infinitesimally slow. With a slow
but finite external driving, there is generally small probability of transition between energy levels.
There has been a great deal of effort to address this small deviation from the quantum
adiabatic theorem~\cite{Rigolin,expand1,expand2,Wuz,continuous,MaamacheM,degeneracy,RigolinGPRL104,TongPRL2005,yukalov,quench,errorbound}.
During the many studies of this issue, a controversy on the validity of the quantum adiabatic theorem arises~\cite{BarryPRL2004,resonant,resonant1,resonant2,real,controversy,controversy1,controversy2,controversy3,controversy4,controversy5,controversy6}.
With the success of the quantum adiabatic algorithm  in quantum computing, this issue has become
also important in a practical sense~\cite{first,robust}. It is hence important to make effort towards a better assessment
and control of the errors in quantum adiabatic computing.

In this work we present a theory to address the deviation from the quantum adiabatic theorem.
Based on a classical mechanics framework, we construct iteratively a hierarchy of Hamiltonians with the zeroth-order Hamiltonian being determined by the original Hamiltonian. The deviations of the $k$th order are the adiabatic invariances
of the $k$th-order Hamiltonian while the adiabaticity of the $k$th-order Hamiltonian is determined by the time derivatives
of the external parameters (denoted $R$) up to the $k$th-order. Within this theoretical framework,
the deviations from the quantum adiabatic theorem can be computed to arbitrary order iteratively.
The theory breaks down at the $k$th-order when the $k$th-order time derivative of the external
parameters  becomes relatively large. We use two simple examples,
the Landau-Zener model  and the spin-1/2
under a rotating magnetic field, to illustrate our hierarchical theory.

\begin{figure}[!htbp]
\vspace*{-0.5cm}
\centering
\includegraphics[width=0.55\textwidth]{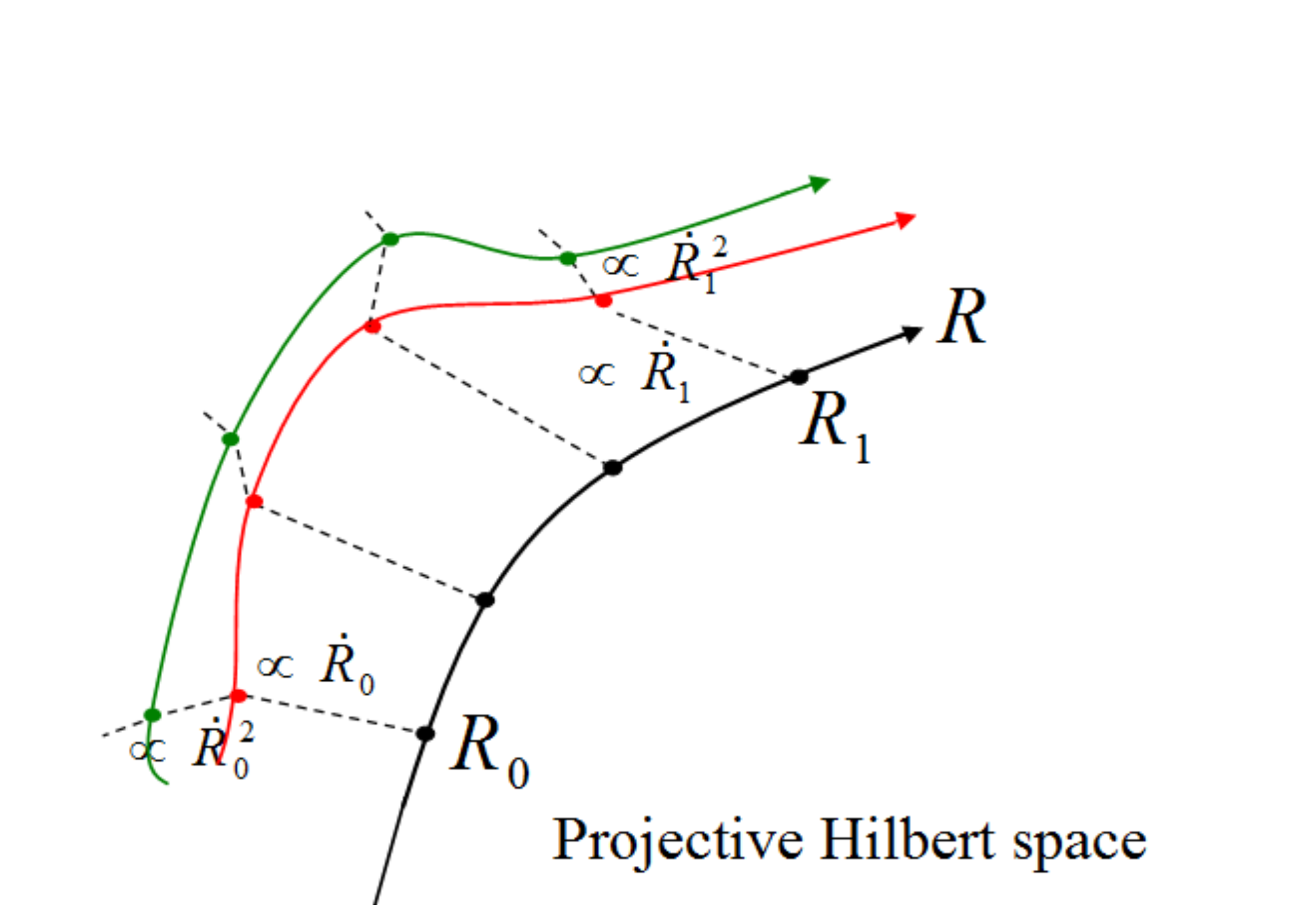}
\caption{(Color online) The adiabatic trajectories of different orders in the projective
Hilbert space. The black line is for the zeroth-order,  the red line for the first-order,
and the dark green line for the second-order.  The difference between the zeroth-order
and the first-order trajectories is proportional to $\dot{R}$ while
the difference between the first-order and the second-order trajectories is proportional
to $\dot{R}^2$.  The possible small oscillations around the adiabatic trajectories
of the first and second orders are omitted for clarity.}
\label{adtr}
\end{figure}

Our hierarchical theory establishes an intuitive picture for quantum adiabatic evolution.
At the zeroth-order, the adiabatic evolution is a smooth curve of instantaneous eigenstates
in the projective Hilbert space where the overall phase is removed.  We call the smooth curve
adiabatic trajectory(see Fig.\ref{adtr}). At the first order,  this adiabatic trajectory is shifted
by a small amount that is proportional to the first time derivative of external parameters
($\dot{R}=dR/dt$).   At the second order, the adiabatic trajectory is shifted again
by a small amount that is proportional to
$\dot{R}^2$ or other possible second-order small parameters, such as  $\ddot{R}$. Predicting and understanding such type of
net shift of a certain order from perfect adiabatic following is one noteworthy feature of our theory.
A schematic picture is presented in Fig.~\ref{adtr}. Depending on the explicit time dependence of $R$,
an actual time evolving state may or may not have small oscillations
around a trajectory that is systematically shifted from the idealized adiabatic trajectory.

Technically we take advantage of two facts to develop our theory. First, we use  the
superposition principle, which allows us to focus on the adiabatic evolution of each
individual energy eigenstate. Second, we use the classical Hamiltonian formulation
of the Schr\"odinger equation~\cite{Weinberg,HeslotPRD1985,Liu2003PRL}. In this formalism,
an energy eigenstate is mapped into an elliptic fixed point in the corresponding projective
Hilbert space. Note that this classical formulation is purely mathematical and is {\it not}
the traditional semiclassical limit $\hbar\rightarrow0$.  Our classical mechanics framework
exposes a deep, previously unknown connection between classical adiabatic theory
and the quantum adiabatic theory.  The relation between classical adiabatic theory
and the quantum adiabatic theory was explored in Ref.~\cite{Liu2003PRL}
but not as deeply as in this work. In particular, a high-order deviation in the
quantum adiabatic following still has a classical mechanics structure and may
be still understood by classical adiabatic theory.

\section{Classical Hamiltonian formulation of the Schr\"odinger equation}
We consider a quantum system described by the Hamiltonian $\hat{H}_0(R)$,
where $R=R(t)$ represents time-dependent parameters in an adiabatic protocol. As normally
assumed for quantum adiabatic evolutions~\cite{adiabatic},  $\hat{H}_0(R)$  has
a discrete non-degenerate spectrum during the entire control protocol.  Further,
the rate of change in $R$ is small as compared with the transition frequencies
of the system.  Deviations from the quantum adiabatic theorem are expected so
long as the protocol is not executed in the mathematical limit $\dot{R} \rightarrow 0$.
The aim of this work is to develop a general and systematic framework to
quantitatively describe such deviations.

Though our consideration can be extended to cases with a Hilbert space of infinite dimensions,
for convenience we assume
$\hat{H}_0(R)$ lives in a finite $n$-dimensional Hilbert space.   $\hat{H}_0(R)$  can thus be expressed as a $R$-dependent $n\times n$ Hermitian matrix.  We find it mathematically more convenient to
use the classical Hamiltonian formulation for  the  Schr\"odinger equation~\cite{Weinberg,HeslotPRD1985,Liu2003PRL}.
We express the quantum state with an $n$-component wavefunction $|\psi\rangle=(c_1,c_2,\ldots,c_n)^T$
and define $n-1$ pairs  of canonical variables
\begin{equation} \label{definition}
p_i=\arg(c_{i+1})-\arg(c_1), \quad q_i=|c_{i+1}|^2,
\end{equation}
with $i=1,2,\ldots,n-1$. By construction,  the Schr\"odinger equation then yields the following Hamilton's equations of motion,
\begin{equation} \label{classical}
\quad \frac{d p_i}{dt}=-\frac{\partial H_0(R)}{\partial q_i}\,,
~~~\frac{d q_i}{dt}=\frac{\partial H_0(R)}{\partial p_i},
\end{equation}
where the classical Hamiltonian $H_0(R)$ is obtained from the quantum Hamiltonian  $\hat{H}_0(R)$ as
\begin{equation} \label{expectation}
H_0(R)=\langle\psi|\hat{H}_0(R)|\psi\rangle\,.
\end{equation}
As the overall phase is removed, the phase space in this classical formalism
is just the projective Hilbert space.
This alternative formalism of the Schr\"{o}dinger equation will allow us to  exploit
powerful and familiar tools in classical mechanics in our analysis.
The overall phase of the wavefunction, or equivalently $\arg(c_1)$, is removed in Eq.~(\ref{classical}) \cite{Weinberg,HeslotPRD1985,Liu2003PRL}

It is particularly interesting to look at eigenstates.  In  the
original Schr\"odinger equation picture, an energy eigenstate of $\hat{H}_0(R)$
at a fixed $R$ simply develops a trivial overall phase.  Since  the
overall phase is discarded in our formalism, such an eigenstate evolution is mapped
to a fixed point in the classical phase space of $H_0(R)$.
The issue of the adiabatic
following with the instantaneous energy eigenstates of $\hat{H}_0[R(t)]$ now
becomes the issue of the adiabatic following with the instantaneous fixed points of $H_0[R(t)]$.

In principle, the time evolution emanating from an arbitrary initial state as a superposition
of different energy eigenstates can be considered. However, the linearity of the original
Schr\"odinger equation indicates that it suffices to study initial states that are
energy eigenstates of $\hat{H}_0[R(0)]$ at $t=0$.
As such, in our classical formalism we only need to consider those initial conditions
that are fixed points in the phase space.

One final technical comment is in order. The mapping from the wavefunction components $c_i$ to phase space variables $(p_i, q_i)$ [see  Eq.~(\ref{definition})] becomes ambiguous when any one of
the wavefunction component $c_i$ becomes zero. Fortunately, this ambiguity can be easily overcome
by adopting  a different representation to re-express the wavefunction.  For example, $c_1$ in Eq.~(\ref{definition})  is used to remove the overall wavefunction phase.  If $c_1=0$,  one can always
select another nonzero $c_i$ to carry out a similar mapping.


\section{First order deviations}
As the generalization to arbitrary dimensions is straightforward,  we consider a
quantum system with a two-dimensional Hilbert space for the rest of the paper.
With $n=2$ the Hamilton's equations of motion in Eq.~(\ref{classical}) only involve
one pair of canonical variables $q_1$ and $p_1$. The phase space is hence also
two-dimensional.  For clarity we drop the subscript $1$ hereafter.  A $R$-dependent
fixed point in the phase space is denoted as $[\bar{p}(R),\bar{q}(R)]$.
There are two fixed points corresponding to two energy eigenstates of $\hat{H}_0(R)$.

According to the quantum adiabatic theorem, under a  sufficiently slow protocol $R=R(t)$,
the dynamics emanating from an energy eigenstate will follow the instantaneous energy
eigenstates. With the removal of the overall phase, this dynamics is completely
described by the smooth curve of instantaneous energy
eigenstates in the projective Hilbert space.  We shall call it  adiabatic trajectory (see Fig. \ref{adtr}).
However, in a realistic protocol
where $R(t)$ changes slowly with a nonzero rate, there should be a deviation
from this picture of perfect adiabatic following.

There were studies on the small deviations from what the adiabatic theorem predicts. It was done
in special classical systems and the small deviations were found to  pollute the Hannay's angle
\cite{Berry1996, adam, mag}. Recently,  the first-order deviation was studied in nonlinear quantum adiabatic evolutions \cite{liu,liu1}, where the result was used successfully to predict a new kind of geometric phase beyond the traditional Berry phase. As their focus was on the global effects of the deviations,
detailed dynamics of the deviation was not considered. Our work
conducts a systematic study of the quantum adiabatic evolution and reveals its hierarchical structure.
Our results can be easily generalized to classical systems and nonlinear quantum systems.

With possible deviations from the instantaneous fixed points $[\bar{q}(R),\bar{p}(R)]$,
the actual adiabatic trajectory in the phase space  can be written as
\begin{equation} \label{expansion}
p(t)=\bar{p}[R(t)]+\delta p,\ \  q(t)=\bar{q}[R(t)]+\delta q,
\end{equation}
with $(\delta p, \delta q)$ being time-dependent deviations from the ideal adiabatic trajectory $[\bar{p}(R),\bar{q}(R)]$.  This section is mainly to develop a theory to understand the behavior of $(\delta p, \delta q)$ to the first order of $\dot{R}$.

As a preparation we first consider the case when $R$ is fixed.
Using Hamilton's equations of motion and Taylor expanding $\frac{\partial H_0(R)}{\partial p}$ and $\frac{\partial H_0(R)}{\partial q}$ to the first order of
 $(\delta p, \delta q)$, we have
\begin{equation} \label{fd}
\left(\begin{array}{c}\frac{dp}{dt}\\ \\
\frac{dq}{dt}\end{array}\right)=\Gamma_0
 \left(\begin{array}{c} \delta p\\ \\
 \delta q\end{array}\right),
\end{equation}
where
\begin{equation} \label{Gamma}
\Gamma_0=\left(\begin{array}{cc}-\frac{\partial^2H_0}{\partial q\partial
p}&-\frac{\partial^2 H_0}{\partial q\partial q}
\\ \\ \frac{\partial^2H_0}{\partial p
\partial p}&\frac{\partial^2H_0}{\partial
p \partial q}
\end{array}\right)_{p=\bar{p},q=\bar{q}}
\end{equation}
is an $R$-dependent matrix obtained from the second-order derivatives of $H_0(R)$.
The terms with first-order derivatives of $H_0(R)$ do not appear on the right-hand side of Eq.~(\ref{fd})  simply because $[\bar{q}(R),\bar{p}(R)]$ is  a fixed point.
All higher-order terms are neglected here.

We now consider the dynamics of $(\delta q, \delta p)$ in the control protocol
where $R=R(t)$ changes slowly with time. In this case, we have
\begin{eqnarray}
\frac{dp}{dt} & = & \frac{\partial \bar{p}(R)}{\partial R} \dot{R} + \frac{d \delta p}{dt} \nonumber \\
\frac{dq}{dt} & = & \frac{\partial \bar{q}(R)}{\partial R} \dot{R} + \frac{d \delta q}{dt}.
\end{eqnarray}
Equation ~(\ref{fd}) consequently becomes
\begin{equation} \label{afd}
\left(\begin{array}{c}\frac{d\delta p}{dt}\\ \\
\frac{d\delta q}{dt}\end{array}\right)=\Gamma_0(R)
 \left[\left(\begin{array}{c} \delta p\\ \\
 \delta q\end{array}\right)-\Gamma_0^{-1}(R)\left(\begin{array}{c}\frac{\partial \bar{p}}{\partial R}\\ \\
 \frac{\partial \bar{q}}{\partial R}\end{array}\right)
 \dot{R} \right].
\end{equation}

Two remarks are necessary for  this equation of $(\delta p, \delta q)$. First,
because it is already assumed that throughout the protocol $R=R(t)$ the studied
energy eigenstates never become degenerate,  the corresponding fixed points
in the phase space do not vanish or collide.   It is therefore legitimate to always
associate the deviations with one fixed point so long as $(\delta p, \delta q)$ is small.
Second, it can be shown that the determinant $|\Gamma_0|$ does not vanish with
non-degenerate energy eigenstates. $\Gamma_0^{-1}$ in Eq.~(\ref{afd}) hence exists for all $R$.

Remarkably, Eq.~(\ref{afd})
possesses a canonical structure. The variables $(\delta p, \delta q)$
are a canonical pair and Eq.~(\ref{afd}) can be derived from  the following Hamiltonian
\begin{eqnarray} \label{Hflu} \nonumber
H_1(R,\dot{R})&=&\frac{1}{2}\left(\frac{\partial^2H_0}{\partial q^2}\right)_{\bar{p},\bar{q}}(\delta q-B_1)^2 \\ \nonumber
&+&\left(\frac{\partial^2H_0}{\partial q\partial p}\right)_{\bar{p},\bar{q}}(\delta q-B_1) (\delta p-A_1) \\
&+&\frac{1}{2}\left(\frac{\partial^2H_0}{\partial p^2}\right)_{\bar{p},\bar{q}}(\delta p-A_1)^2,
\end{eqnarray}
where $A_1=A_1(R,\dot{R})$ and $B_1=B_1(R,\dot{R})$ are defined as
\begin{equation} \label{center}
\left(\begin{array}{c}A_1\\ \\
B_1\end{array}\right)=\Gamma_0^{-1}(R)\left(\begin{array}{c}\frac{\partial \bar{p}}{\partial R}\\ \\
 \frac{\partial \bar{q}}{\partial R}\end{array}\right) \dot{R}\,.
\end{equation}
This expression was previously obtained by Fu and Liu \cite{liu,liu1}.
It is clear that the first-order Hamiltonian (\ref{Hflu}) describes harmonic
oscillations around  the central point $(A_1,B_1)$.

The first-order  Hamiltonian $H_1$ generating the dynamics of $(\delta p, \delta q)$
depends upon two parameters $R(t)$ and $\dot{R}(t)$. We assume that $\dot{R}(t)$ also
changes slowly with time. In this case, the dynamics of $(\delta p, \delta q)$
becomes the adiabatic evolution of $H_1$ and can be understood with the help of the classical
adiabatic theorem. We define the action for $(\delta p,\delta q)$ as
\begin{equation} \label{de}
I_1=\frac{1}{2\pi}\oint \delta p\cdot d(\delta q)\,.
\end{equation}
This action is the adiabatic invariant  possessed by $H_1$~\cite{cla-adia}.
$(A_1,B_1)$ is the fixed point of $H_1$ with $I_1=0$.  The dynamics of $(\delta p,\delta q)$
can be viewed as a spiral motion along the adiabatic trajectory specified by fixed point $(A_1,B_1)$.
The amplitude of the spiral oscillations is determined by the action $I_1$. With this analysis,
it becomes clear that  when both $R(t)$ and $\dot{R}(t)$ change slowly with time
$(A_1,B_1)$ describes an adiabatic trajectory shifted from
the ideal trajectory of fixed point $[\bar{p}(R), \bar{q}(R)]$ as shown in Fig.~\ref{adtr}.

\begin{figure}[!htbp]
\centering
\includegraphics[width=0.60\textwidth]{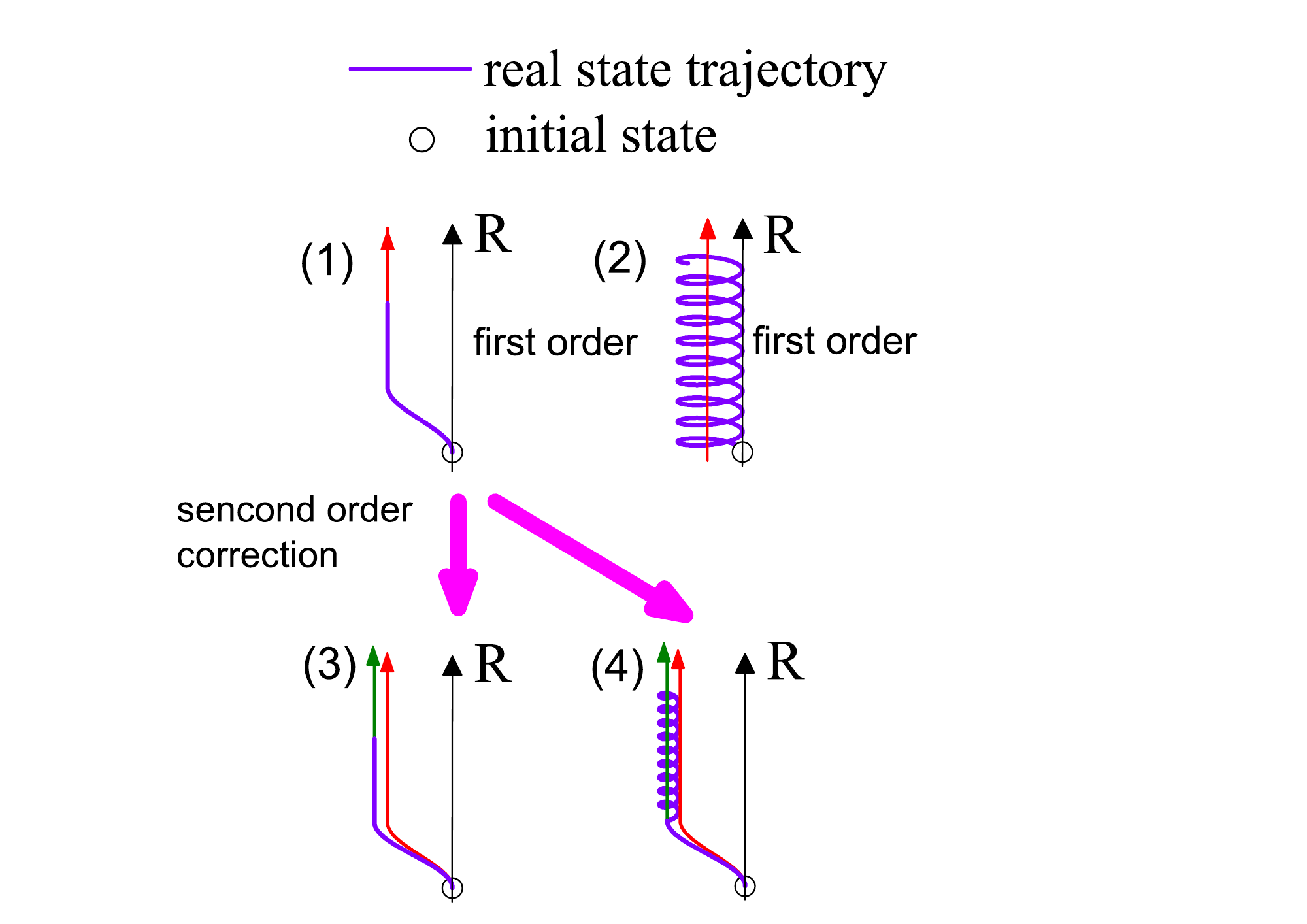}
\vspace*{0.2cm}
\caption{(Color online) Adiabatic evolutions at first and second orders. The black line is the zeroth-order adiabatic trajectory,  the red line the first-order adiabatic trajectory, and the dark green line the second-order adiabatic trajectory.
(a) The evolution follows the first-order adiabatic trajectory when the adiabatic
manipulation is gradually launched (continuous increasing of $\dot{R}$ from zero);
(b) it becomes a spiral oscillatory motion when the process is started with a finite $\dot{R}$.
(c) The state follows  the second-order adiabatic trajectory when $\ddot{R}$ is
changed slowly from zero; (d) it becomes a spiral-like motion when $\ddot{R}$ is started with
a finite value.}
\label{adtr22}
\end{figure}

We now consider two typical cases. In the first case, $\dot{R}$ is increased
slowly from zero. In this case, as $A_1$ and $B_1$ are zero initially,
the action $I_1$ is zero and the adiabatic evolution to the first order
will follow exactly the adiabatic trajectory specified  by $(A_1,B_1)$. This is
illustrated in Fig.~\ref{adtr22}(a).  In the second case,  the external driving rate
$\dot{R}$ is finite and small at the beginning. This means that $(A_1,B_1)$
is not zero initially and the action $I_1$ has a finite and small value. In this second case,
the adiabatic evolution will become a spiral motion around the
trajectory of $(A_1,B_1)$ as shown in Fig.~\ref{adtr22}(b).
This analysis of the second case in fact implies that infrequent sudden but small
jump of $\dot{R}$ will not break down the adiabaticity of the evolution.
Note that the smallness of the jump in $\dot{R}(t)$ is implicitly guaranteed
by the slow change of $R(t)$. We mention it explicitly in our discussion just for clarity.

Our first-order adiabatic theory  shows that
a small quantum transition to other energy eigenstates always occurs with probability
proportional to $\dot{R}$. The probability is zero only in special cases where  the coefficients in Eq.~(\ref{center}) vanish.


Our first-order theory offers a  deep insight into the generic subtlety of how the adiabatic following
breaks down. Let us consider a situation where $\dot{R}(t)$ is small but changes with a great rate, i.e.,
$\ddot{R}(t)$ is large. In this case, the dynamics governed by $H_1$ is not adiabatic; $I_1$ is not an adiabatic invariant and can not stay small for a long time. When the evolution is long enough,
the dynamical evolution of the first-order deviations $(\delta p, \delta q)$ will no longer be bounded:
the small deviations $(\delta p, \delta q)$ can accumulate and eventually be amplified to the
zeroth-order level.  This breakdown due to the largeness of $\ddot{R}(t)$ clearly depends
on the detail of the protocol  $R(t)$ and the Hamiltonian; general conclusions will
be difficult to reach.

We note that our theory can be naturally extended to a Hilbert space of larger dimension $n>2$,
where the matrix $\Gamma_0$ becomes $2(n-1) \times 2(n-1)$ dimension and the first-order
Hamiltonian has $(n-1)$ pairs of canonical variables.

\section{Second order deviations}
In the previous section  we have found that the first-order correction $(\delta p, \delta q)$ evolves
according to a first-order Hamiltonian $H_1$. It is natural to wonder whether we can find a similar
Hamiltonian for the second-order deviations. We find that if the system follows the first-order
adiabatic trajectory (see Fig.~\ref{adtr22}(a)),  we can indeed find such a Hamiltonian.
We write
\begin{equation}
p=\bar{p}+A_1+\delta^2p\,,~~~~~~q=\bar{q}+B_1+\delta^2q\,.
\end{equation}
After substituting it into $H_0$ and with straightforward calculation,  we obtain the second-order
Hamiltonian
\begin{eqnarray} \label{2ndH}
\nonumber
&&H_2(R, \dot{R}, \ddot{R})=\frac{1}{2}\left(\frac{\partial^2 H_0}
{\partial q\partial q }\right)_{\bar{p}+A_1,\bar{q}+B_1}(\delta^2 q-B_2)^2 \\ \nonumber
&&+\left(\frac{\partial^2H_0}{\partial q\partial p}\right)_{\bar{p}+A_1,\bar{q}+B_1}(\delta^2 q-B_2) (\delta^2 p-A_2)\\&&+\frac{1}{2}\left(\frac{\partial^2H_0}{\partial p \partial p}\right)_{\bar{p}+A_1,\bar{q}+B_1}(\delta^2 p-A_2)^2\,,
\end{eqnarray}
where
\begin{eqnarray} \label{center2z}  \nonumber
\left(\begin{array}{c} A_2\\ \\
 B_2\end{array}\right)&=& \Gamma_0^{-1}\left[\left(\begin{array}{c}
 \frac{\partial A_1}{\partial R}\\ \\
\frac{\partial B_1}{\partial R}\end{array}\right)
\dot{R}+\left(\begin{array}{c}\frac{\partial A_1}{\partial \dot{R}}\\ \\
\frac{\partial B_1}{\partial \dot{R}}\end{array}\right)
\ddot{R}\right] \\  &-&\frac{1}{2}\Gamma_0^{-1} \delta\Gamma \left(\begin{array}{c} A_1\\ \\
 B_1\end{array}\right)
 \,.
\end{eqnarray}
Here  $\delta\Gamma$ is defined as
\begin{equation} \label{shiftgamma}
\delta\Gamma=\left(\frac{\partial\Gamma}{\partial p}\right)_{\bar{p},\bar{q}}A_1+\left(\frac{\partial\Gamma}{\partial q}\right)_{\bar{p},\bar{q}}B_1
 \end{equation}
with
\begin{equation} \label{Gamma}
\Gamma\equiv \left(\begin{array}{cc}-\frac{\partial^2H_0}{\partial q\partial
p}&-\frac{\partial^2 H_0}{\partial q\partial q}
\\ \\ \frac{\partial^2H_0}{\partial p
\partial p}&\frac{\partial^2H_0}{\partial
p \partial q}
\end{array}\right)\, .
\end{equation}
The  detailed derivation of this second-order Hamiltonian (\ref{2ndH}) can be found in Appendix A
along with some subtlety involved in the derivation.

The second-order  Hamiltonian $H_2$ has a similar structure as $H_1$ and describes
a generalized harmonic oscillator. The significant difference is that $H_2$ depends on three parameters
$(R, \dot{R}, \ddot{R})$ while $H_1$ depends on only two parameters $(R, \dot{R})$.
In the following, we conduct a similar analysis for  $H_2$ as for $H_1$.
We focus on the case where $\ddot{R}$, along with $R, \dot{R}$, changes slowly with time.
In this case, the dynamics of the second-order deviation $(\delta^2p, \delta^2q)$ as governed by $H_2$
is adiabatic.  We define the action for $(\delta^2p,\delta^2q)$ as
\begin{equation} \label{de2}
I_2=\frac{1}{2\pi}\oint \delta^2 p\cdot d(\delta^2 q)\,,
\end{equation}
which is the adiabatic invariant possessed by $H_2$~\cite{cla-adia}.
$(A_2,B_2)$ is the fixed point of $H_2$ with $I_2=0$.  The dynamics of $(\delta^2 p,\delta^2 q)$
can be viewed as a spiral motion along the adiabatic trajectory specified by fixed point $(A_2,B_2)$.
The amplitude of the spiral oscillations is determined by the action $I_2$.
It is clear from this analysis  that  $(A_2,B_2)$ describes an adiabatic trajectory shifted from
the first-order  one that is specified by $[\bar{p}+A_1, \bar{q}+B_1]$ (see Fig.~\ref{adtr}).

We again consider two typical cases. ({\it i}) When both $\dot{R}$ and $\ddot{R}$ are started
continuous from zero, $I_2$ is zero and the dynamics of $(\delta^2p,\delta^2q)$ follows exactly $(A_2,B_2)$. This means that the state follows exactly the adiabatic trajectory deviating from original instantaneous
eigenstate by $(A_1+A_2,B_1+B_2)$ (see Fig.~\ref{adtr22}(c)).  ({\it ii}) When the system starts with
a finite $\ddot{R}$, $I_2$ is nonzero and the system undergoes a spiral motion around $(A_2,B_2)$ (see Fig.~\ref{adtr22}(d)). The amplitude of the spiral motion is determined by $I_2$.

We can continue this procedure and construct a $k$th-order Hamiltonian for the $k$th-order deviation.
The result and the detailed derivation can be found in Appendix B. A general feature is that
the $k$th-order Hamiltonian will depend on $k+1$ parameters,
$R, \dot{R}, \ddot{R}, \cdots, d^{k} R/dt^{k}$, and the adiabaticity of its dynamics is controlled
by these parameters.  We note that a $k$th-order Hamiltonian can be constructed only when
the dynamics of the deviations of order $(k-1)$ follows the $(k-1)$th-order adiabatic trajectory
(the scenarios illustrated in Fig.~\ref{adtr22}(a,c)).



In brief, we have developed a hierarchical theory for quantum adiabatic evolution. In this theory,
a hierarchy of Hamiltonians can be constructed: the $k$th-order deviation from quantum adiabatic theorem
is governed by a $k$th-order Hamiltonian. This theory not only offers explicit formula to compute
the deviations of various orders but also presents an intuitive insight into the intricacy of adiabatic evolution.
To illustrate the latter, we use the second-order Hamiltonian $H_2(R,\dot{R},\ddot{R})$ as an example.
We assume that $R,\dot{R}$ is small while $\ddot{R}$ is large. In this case
the dynamics of the second-order deviation
$(\delta^2p,\delta^2q)$  governed by $H_2$ is not adiabatic. As a result,
the second-order deviation $(\delta^2p,\delta^2q)$ can grow, reach
the first-order level, and continue to grow even bigger. The evolution of the first-order deviations
is adiabatic due to the smallness of $R$ and $\dot{R}$. However, this conclusion is only true
when the  deviation is small. If the second-order deviation grows so large that the deviation is no longer
small, the adiabticity at the first-order level is then broken. Eventually, the growth starting from
the second-order level  can even break down the zeroth-order adiabaticity.
This example suggests that the adiabatic evolution can be maintained for an arbitrary long time
only when all orders of time derivative of $R$ are small. However,
such growth of a high-order deviation to an appreciable quantity at a low-order level
can take a long time scale beyond our practical interest.  As the exact time scale needed for this growth
depends on the detail of the control protocol $R(t)$, it can only be examined case by case.
Finally, as we discuss below,  a breakdown of adiabaticity at a higher-order may not pass on to a lower-level and then cause
the breakdown of adiabaticity at the lower level.

\begin{figure}[t]
\vspace*{-0.5cm}
\centering
\includegraphics[width=0.6\textwidth]{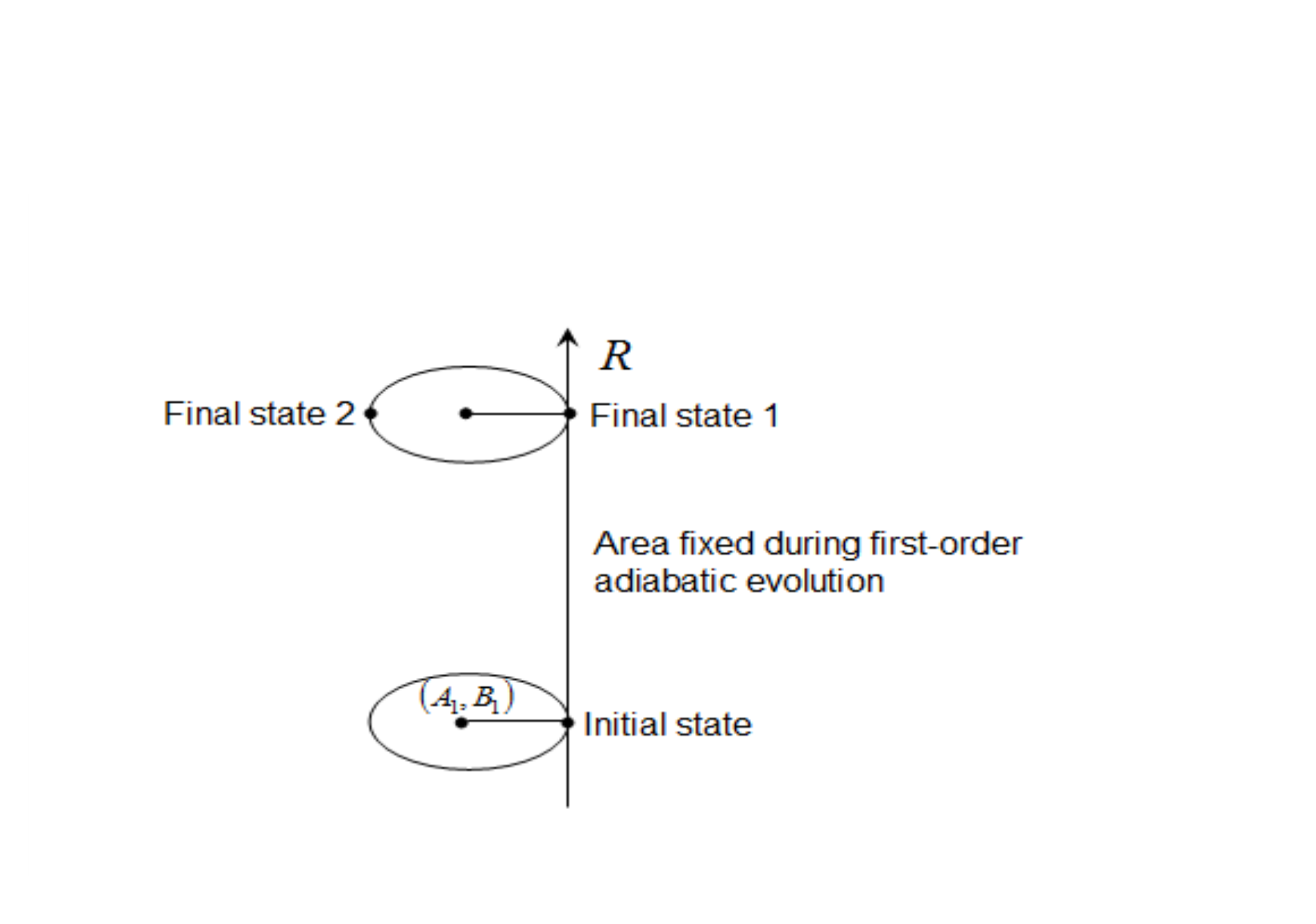}
\vspace*{-1.0cm}
\caption{Illustration of initial and final states due to spiral-like motion, where the initial state is exactly the instantaneous eigenstate and, at the initial and final times, the spiral-like motions are all in tangency with the trajectory of the instantaneous eigenstate. The mean first-order deviations $(A_1,B_1)$ are not zero on the starting and ending points of the adiabatic process and the area enclosed by orbit, or $I_1$, determined by $(A_1,B_1)$ on the starting point is conserved according to the first-order adiabatic theory. The orbits here in the first-order approximation permit second-order errors.}
\label{indication}
\end{figure}

In Fig.~\ref{indication}, we illustrate a specific case that, for $(A_1,B_1)\neq(0,0)$, the spiral-like motions are in tangency with the trajectory of the instantaneous eigenstate on both the starting and ending times of the adiabatic process. If the duration of the adiabatic process is precisely chosen such that the final state is just on the instantaneous eigenstate, then the adiabaticity is accidently restored, a situation different from the true adiabaticity maintained throughout the whole process. Because the adiabatic curve itself is derived within the first-order approximation, to the second-order accuracy the final state is not exactly on the instantaneous eigenstate (see point 1 in Fig.~\ref{indication}). On the other hand, if the duration $T$ is chosen such that the final state is on point 2 (see Fig.~\ref{indication}), then the final state will deviate from the instantaneous eigenstate with a first-order deviation. For a higher-order case, e.g., $(A_1,B_1)=(0,0)$ but $(A_2,B_2)\neq(0,0)$ at the starting and ending times of the protocol,
we will have an analogous situation. These qualitative insights are fully consistent with early results in Ref.~\cite{higher-order-domi1} based on a different approach.

We can now also see the possibility of higher-order deviations not accumulating to a lower-order deviation from the perfect adiabatic following.  Suppose we divide the whole protocol into many segments.  If, at the end of each segment, the state rotates (with $k$th order spiral-like motion when $(A_i,B_i)=(0,0)$ for $i=1,2,\ldots k-1$ but $(A_k,B_k)\neq(0,0)$) back to the instantaneous eigenstate, then this $k$th order deviation is unable to accumulate to the $(k-1)$th order. By contrast, if the state at the ending times of many segments always rotates away, say, to the farthest point from the instantaneous eigenstate, then the deviation can become larger and larger and eventually its value may accumulate to reach the $(k-1)$th order.


For the hierarchical expressions of adiabatic errors detailed in Appendix B, we have assumed that the higher time derivatives of $R$ possess a higher-order (and hence smaller) magnitude, e.g., the term proportional to $(d^mR/dt^m)^n$ belongs to the group of terms of $m\cdot n$-order.  This grouping scheme, mainly for convenience, is intuitive and can be reasonable in a vast variety of adiabatic protocols.  However, this order grouping scheme may be problematic in some protocols of $R(t)$. Consider, for example, the protocol $R=\sin(\epsilon t)$ with a small $\epsilon$, then all the higher-order derivatives will be small, and $(d^mR/dt^m)^n\propto \epsilon^{mn}$ is indeed a $m\cdot n$-order term. However, for the protocol $R=\epsilon\sin(t)$ with a small $\epsilon$,  all orders of time derivatives of $R$, e.g., $\dot{R}, \ddot{R},\ldots,d^kR/dt^k$, are of the same order of magnitude. As such, for this situation a term containing $(d^mR/dt^m)^n\propto \epsilon^{n}$ is of the $n$-order. On the one hand, this is fully consistent with early observations that sometimes terms associated with higher-order derivatives of $R$ can be important \cite{higher-order-domi1,higher-order-domi2}. On the other hand,  it is clear now that the many terms arising from our hierarchical theory may not automatically be an expansion cast in terms of their orders of magnitude.  To analyze the details we still need to make use of the explicit $R(t)$ to assess the actual importance (or weightage) of the many different terms emerging from our theory.  In any case, it is learnt from our classical mechanics framework that the dynamics of quantum adiabatic following can be digested in terms of adiabatic following of various orders occurring in parallel.

\section{Two examples}
We now use two simple systems to illustrate our hierarchical theory. One is a spin-1/2
particle in an external rotating magnetic field; the other is the Landau-Zener model. They are
chosen because they are either exactly solvable or their numerical solutions can be found
with great accuracy.  In this way, there will be no ambiguity in
checking the validity of our hierarchical theory. In this section, we always assume $\hbar=1$.

\subsection{spin-1/2 under a rotating field}

In the hierarchical theory, the first-order deviation and its dynamics is of the most importance. In this subsection, we employ the simple model of a spin-1/2 particle in a rotating magnetic field to
illustrate the first-order adiabatic theory. The Hamiltonian for a spin-1/2 particle in an external rotating field is
\begin{equation} \label{HRo}
\hat{H}_0=\frac{1}{2}\left(\begin{array}{cc}0&L\exp(-i\alpha)
\\L\exp(i\alpha)&0
\end{array}\right),
\end{equation}
where $\alpha(t)$ changes slowly with time for a rotating field. We use
$|\psi\rangle=(c_1,c_2)^T$,  where $c_1$ and $c_2$ are complex, to
denote the quantum state of this spin-1/2 particle. We turn to the
classical formulation by introducing  a pair of conjugate variables,
$p=\arg(c_2)-\arg(c_1)$ and $q=|c_2|^2$.  The corresponding
classical Hamiltonian is
\begin{equation} \label{HRoC}
H_0=\langle\psi|\hat{H}_0|\psi\rangle =L\sqrt{q-q^2}\cos(\alpha-p).
\end{equation}
The classical Hamiltonian in Eq.~(\ref{HRoC}) has two elliptic  fixed points, namely,
$(\bar{q}=1/2,\bar{p}=\alpha)$ and $(\bar{q}=1/2,\bar{p}=\alpha+\pi)$,  corresponding respectively to
the two eigenstates of Eq.~(\ref{HRo}). We focus on the adiabatic following
of the fixed point $\bar{q}=1/2,\bar{p}=\alpha$ as $\alpha$ (rotating field) changes slowly.
The conventional adiabatic theorem states that the actual state will accurately follow
the instantaneous state $(\bar{q}=1/2,\bar{p}=\alpha)$.

On top of the conventional adiabatic theorem, there are first-order corrections. To that end we now derive the effective first-order Hamiltonian $H_1$. According to Eqs.~(\ref{Hflu}), (\ref{center}) and (\ref{HRoC}), one finds for fixed point ($q=1/2,p=\alpha$),
\begin{equation} \label{Hflu1}
H_1=-\frac{L}{2}\left[2(\delta q-\frac{\dot{\alpha}}{2L})^2
+\frac{1}{2}(\delta p)^2 \right]\,.
\end{equation}
Interestingly, for this example,  $H_1$ happens to be independent
of the adiabatic parameter $\alpha$.
The first-order fixed point is located at $A_1=0$, $B_1=\dot{\alpha}/2L$.   In the following we consider
three different control protocols $\alpha(t)$ with $\alpha(0)=0$ and the initial state emanating exactly from the fixed point $q(0)=1/2,p(0)=\alpha$.

\begin{figure}[t]
\vspace*{-0.5cm}
\centering
\includegraphics[width=0.45\textwidth]{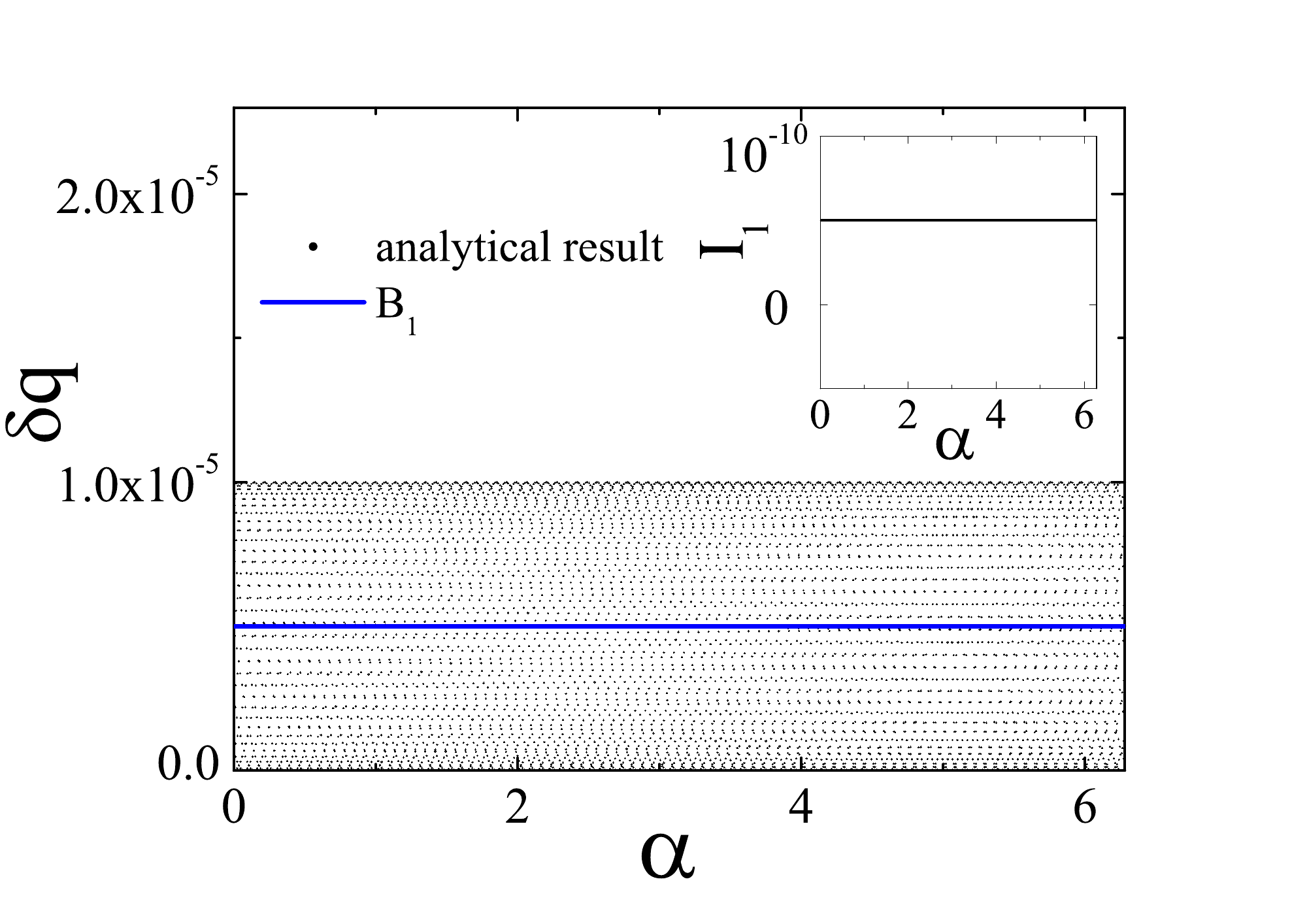}
\caption{The first-order solution of a spin-1/2 particle in a slowly rotating magnetic field.
The dots are $\delta q$ computed from the analytical solution to the first order Eq.~(\ref{analy}).
The solid line is $B_1$ of the first-order fixed point.  The inset shows the first-order action $I_1$
computed from the analytical solution Eq.~(\ref{analy}).
$\omega=10^{-5}$.  }
\label{vt}
\end{figure}

({\it i}) Let us first consider the simplest protocol in which $\alpha=\omega t$ with $\omega$ being constant.
At $t=0$, the initial state is $q=1/2,p=0$ while the first-order fixed point is at $(A_1=0, B_1=\omega/2L)$.
So, the state starts off the first-order fixed point and the first-order action is
\begin{equation} \label{deRo}
I_1=\frac{\omega^2}{4L^2}\,.
\end{equation}
According to our theory, the first-order deviation will undergo a spiral motion, similar to what
is depicted in Fig.~\ref{adtr22}(b), with its amplitude determined by $I_1$.

The validity of our theory can be checked by directly integrating the Schr\"odinger equation governed by (\ref{HRo}). This solution can be found exactly.  With the omission of higher orders, the solution
can be written as
\begin{equation} \label{analy}
|\psi\rangle=\frac{1}{\sqrt{2}}\left(\begin{array}{c}1-\frac{\omega}{2L}\left[1-\cos(Lt)\right]
\\\left[1+\frac{\omega}{2L}(1-\cos(Lt))\right]e^{i[\omega t+\frac{\omega}{L}\sin(Lt)]}
\end{array}\right)\,.
\end{equation}
This solution is plotted in Fig.~\ref{vt} by mapping $|\psi\rangle$ to $(p,q)$ and thus to $(\delta p,\delta q)$.
In this figure, we clearly see oscillations around the fixed point $(A_1=0, B_1=\omega/2L)$, consistent with
our first-order theory. As shown in the inset of this figure,  our direct computation also confirms that
the first-order action $I_1$ is a constant. We point out that this is equivalent to a system
under the following control protocol
\begin{equation}
\alpha=0~~~{\rm for}~~~t<0\,;~~~~\alpha=\omega t~~~{\rm for}~~~t>0\,.
\end{equation}
That is, there can be a small sudden jump in $\dot{\alpha}$  at $t=0$. Analytically,
the first-order deviation can be readily computed from the solution (\ref{analy})
\begin{eqnarray} \nonumber
&&\delta p=p-\bar{p}=\frac{\omega}{L}\sin(Lt), \\
&&\delta q=q-\bar{q}=\frac{\omega}{2L}\left(1-\cos(Lt)\right),
\end{eqnarray}
which is indeed consistent with the first-order Hamiltonian dynamics predicted by $H_1$ in Eq.~(\ref{Hflu1}).

According to the mapping (\ref{definition}) between $\delta p$ and $\delta q$ and wavefunction, we can write down the adiabatic error during the whole adiabatic process in terms of the quantum state,
\begin{equation}
\text{Err}=1-|\langle\psi\left(\bar{p},\bar{q}\right)\mid\psi\left(\bar{p}+\delta p,\bar{q}+\delta q\right)\rangle|^2=\frac{\dot{\alpha}^2}{4L^2}\propto \dot{\alpha}^2,
\end{equation}
which is consistent with an earlier result based on exact calculations \cite{Bohm}.

\begin{figure}[t]
\vspace*{-0.5cm}
\centering
\includegraphics[width=0.45\textwidth]{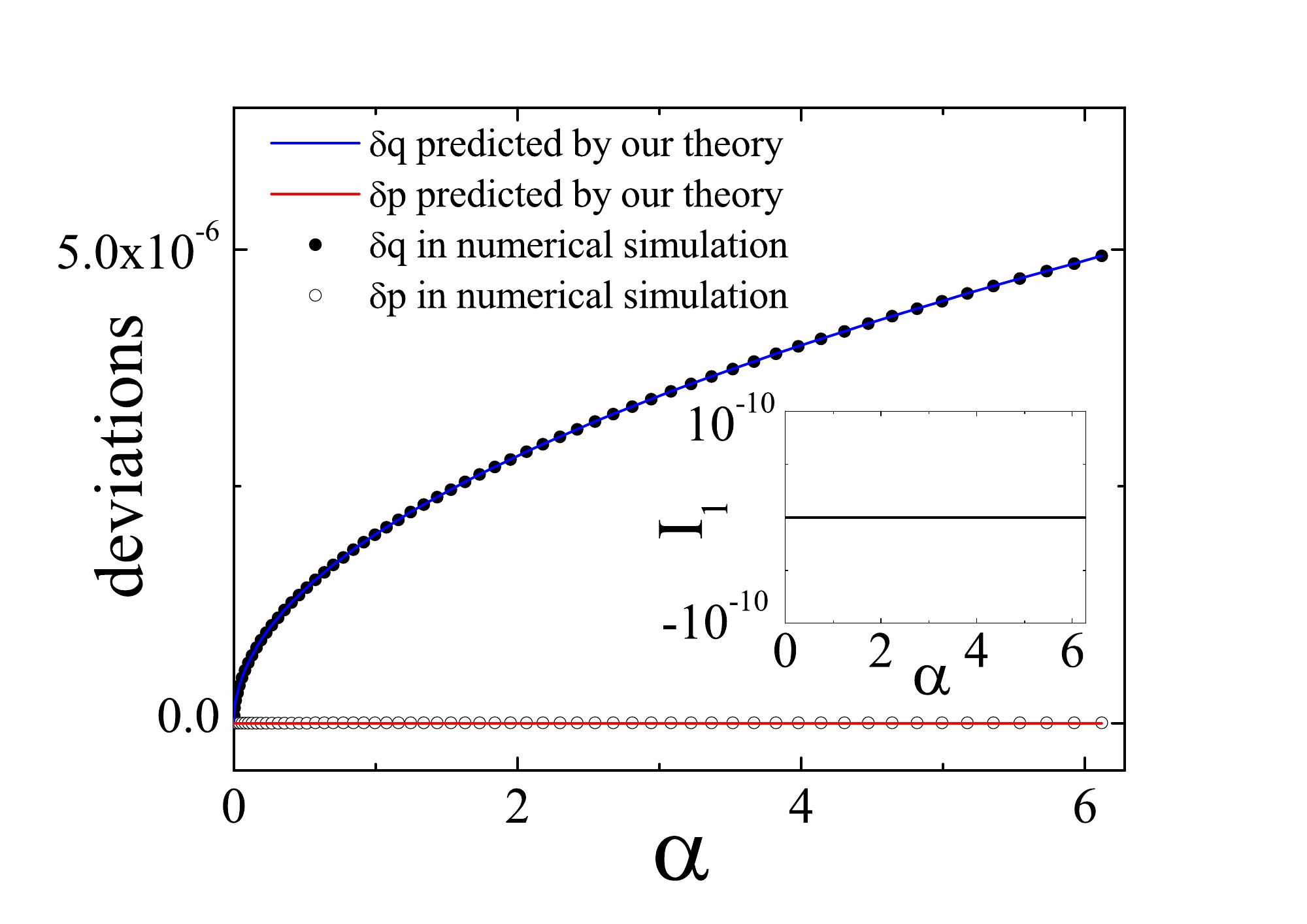}
\caption{Numerical results for the first-order correction $(\delta p, \delta q)$ obtained from
the Hamilton's equation of motion governed by (\ref{HRoC}). The control protocol is
$\alpha=\frac{1}{2}at^2$  with $a=7.96\times10^{-12}$.  The dots and circles are numerical results
while the solid lines are for the first-order fixed point $(A_1=0,B_1=at/2L)$.  The inset shows
the numerically computed $I_1$. }
\label{at2}
\end{figure}

({\it ii}) In the second protocol,  the speed $\dot{\alpha}$ increases gradually from zero.
To be specific, we choose $\alpha=\frac{1}{2}at^2$ with $a=7.96\times10^{-12}$.
For this protocol,  the first-order fixed point is $(A_1=0,B_1=0)$ at $t=0$. Therefore, according to
our first-order theory, the action $I_1=0$ and the dynamics of the first-order deviation
$(\delta p,\delta q)$ follows exactly the first-order fixed point $(A_1=0,B_1=at/2L)$.
We have numerical solved the Hamilton's equations of motion governed by Eq.~(\ref{HRoC})
for this second protocol.  The numerical results for $(\delta p,\delta q)$ and $I_1$ are shown
in Fig.~\ref{at2} and an excellent agreement with our first-order theory is found.

\begin{figure}[t]
\vspace*{-0.5cm}
\centering
\includegraphics[width=0.45\textwidth]{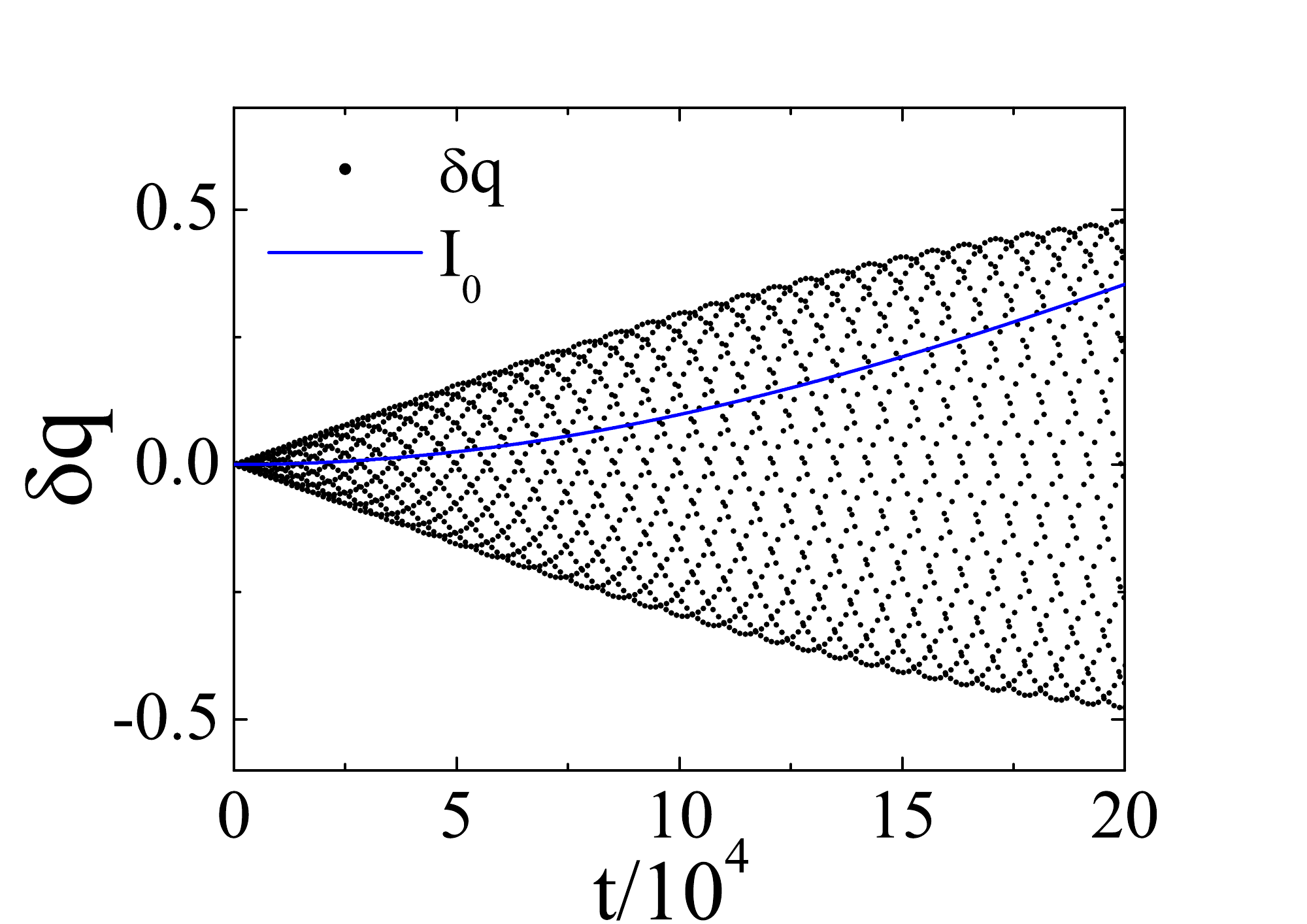}
\caption{Numerical solution of $\delta q$ and $I_0=\frac{1}{2\pi}\oint p\cdot d q$ with Hamilton's equation of motion governed by (\ref{HRoC}). For the control protocol, we have $|\dot{\alpha}|=10^{-5}$. The sign of $\dot{\alpha}$
oscillates with the frequency $\nu=1$.}
\label{rand}
\end{figure}

({\it iii}) In the third protocol, we change the sign of $\dot{\alpha}$ frequently while keeping
$|\dot{\alpha}|$ small. This is to ensure that the second-order time derivative $\ddot{\alpha}$ can
be quite large. We use this protocol to illustrate an insight offered by our hierarchical theory:
high-order time derivative of $R$ can also lead to the breakdown of adiabaticity.
For this spin-1/2 system,  the smallness of $|\dot{\alpha}|$ does not guarantee the accuracy of
the quantum adiabatic theorem.  When $\ddot{\alpha}$ is large, then the first-order
dynamics governed by $H_1$ is no longer adiabatic, and the accumulation of $(\delta p, \delta q)$
will eventually lead to the breakdown of adiabaticity at the zeroth orders.
We have solved numerically the  equations of motion governed by Eq.~(\ref{HRoC}).
The results are plotted in Fig.~\ref{rand}, where we see that $\delta q$ can indeed grow and destroy
the adiabaticity. The solid line seen in the middle of the pattern shown in Fig.~\ref{rand} demonstrates that the action $I_0$ is no longer a constant.

\subsection{Hierarchy of adiabatic corrections in the Landau-Zener model}

In this  subsection, we consider a different model, the Landau-Zener (LZ) model, and use it
to demonstrate  higher-order deviations. The LZ Hamiltonian can be written as
\begin{equation} \label{Hamiltonian-LZ}
\hat{H}_0^{\text{LZ}}=\frac{1}{2}\left(\begin{array}{cc}z&x
\\x&-z
\end{array}\right),
\end{equation}
where the  coupling term $x>0$ is a constant whereas $z$ changes slowly and linearly
from $-Z_0$ to $Z_0$,
\begin{equation}
z=Vt, \quad\quad    t:-Z_0/V\rightarrow Z_0/V.
\end{equation}
Similarly, we define $c_1=|c_1| e^{i\phi_{c_1}}$, $c_2=|c_2| e^{i\phi_{c_2}}$, $p=\phi_{c_2}-\phi_{c_1}$, and $q=|c_2|^2$, and obtain the classical Hamiltonian (drop a constant):
\begin{equation} \label{HamiltonianCLZ}
H_0=\langle\psi|\hat{H}_0^{\text{LZ}}|\psi\rangle=x\sqrt{q-q^2}\cos(p)-zq.
\end{equation}
 This classical Hamiltonian has two fixed points at  $(\bar{p}=0,\bar{q}=\frac{1}{2}-\frac{z}{2\sqrt{x^2+z^2}})$
 and $(\bar{p}=\pi,\bar{q}=\frac{1}{2}+\frac{z}{2\sqrt{x^2+z^2}})$.
Without loss of generality, we focus on the fixed point $(\bar{p}=\pi,\bar{q}=\frac{1}{2}+\frac{z}{2\sqrt{x^2+z^2}})$, which corresponds to the eigenstate with the lower energy.  According to the quantum adiabatic theorem, i.e.,
zeroth-order theory,  when the initial state is the ground state $\left[\cos(\frac{\arctan(x/Z_0)}{2}),-\sin(\frac{\arctan(x/Z_0)}{2})\right]^T$ at $z=-Z_0$  the system will follow the instantaneous eigenstate and ultimately reach $\left[-\sin(\frac{\arctan(x/Z_0)}{2}),\cos(\frac{\arctan(x/Z_0)}{2})\right]^T$ at $z=Z_0$. In what follows, we will compute explicitly  the first-order deviation and the second-order deviation, and
discuss some general properties of the higher-order deviations.

According to Eqs.~(\ref{Hflu},\ref{HamiltonianCLZ}), the first-order Hamiltonian $H_1$ reads
\begin{equation} \label{HLZflu}
H_1=\sqrt{x^2+z^2}\left(1+\frac{z^2}{x^2}\right)(\delta q)^2+
\frac{1}{4}x^2\frac{(\delta p-\frac{V}{x^2+z^2})^2}{\sqrt{x^2+z^2}}.
\end{equation}
The  fixed point for the first-order deviation $(\delta p, \delta q)$ (or the first-order deviation from
the zeroth-order adiabatic trajectory) is
\begin{equation}
\left(\begin{array}{c} A_1\\ \\
B_1\end{array}\right)=\left(\begin{array}{c} \frac{V}{x^2+z^2}\\ \\
0\end{array}\right).
\end{equation}

The results for the second-order deviation $(\delta^2 p, \delta^2 q)$ can be computed
similarly. The second-order Hamiltonian is
\begin{eqnarray} \nonumber
H_2&=&\sqrt{x^2+z^2}\left(1+\frac{z^2}{x^2}\right)\left(\delta^2 q-\frac{5x^2zV^2}{4(x^2+z^2)^{7/2}}\right)^2 \\ \nonumber
&-&\frac{zV}{x^2+z^2}\left(\delta^2 q-\frac{5x^2zV^2}{4(x^2+z^2)^{7/2}}\right)\delta^2 p \\
&+&\frac{1}{4}x^2\frac{(\delta^2 p)^2}{\sqrt{x^2+z^2}}.
\end{eqnarray}
The fixed point (or, the deviation from the first-order adiabatic trajectory) is
\begin{eqnarray} \label{center2LZ} \nonumber
\left(\begin{array}{c} A_2\\ \\
B_2\end{array}\right)&=&\left(\begin{array}{c} 0\\ \\
\frac{5x^2zV^2}{4(x^2+z^2)^{7/2}}\end{array}\right);
\end{eqnarray}

We  consider the limit $Z_0\rightarrow \infty$. At this limit, we have $A_1=B_1=A_2=B_2=0$ at
$|z|=\infty$. This means that the deviations of the first-order and the second-order are zero both
at the beginning and at the end of the evolution. The higher-order deviations can also be computed
with the formula in the Method section. There is no need to write them down here. We only want to mention,
for all these higher-order deviations, we also have
\begin{eqnarray}  \label{LZbegin}
A_k \rightarrow 0; \quad  B_k \rightarrow 0, \quad \text{as}\quad |z|\rightarrow \infty.
\end{eqnarray}
This indicates that the LZ tunneling rate at $Z_0\rightarrow\infty$   tends to zero to all orders of the small
driving rate $V$ based on our hierarchy theory.
This is perfectly consistent with the standard rigorous result for the LZ tunneling rate $\exp(-\frac{\pi x^2}{V})$ \cite{LZ,LZ1}, where any term in the Taylor expansion of $\exp(-\frac{\pi x^2}{V})$
with respect to $V$ is zero. This result is in sharp contrast to the previous case in the last subsection, where the leading term of the deviation from an ideal adiabatic behavior is proportional to $\omega^2$.

\vspace{0.3cm}

\section{Summary}

Our hierarchical theory is summarized in Tab.~1. We have found that the
small deviations from the  quantum adiabatic theorem can be analyzed in a
hierarchical order (not necessarily in terms of the order of magnitude of each term):
the deviations of $k$th-order are governed by
a $k$th-order Hamiltonian which depends on $R,\dot{R},\ldots,d^kR/dt^k$.
When there is a large change in $d^kR/dt^k$, the dynamics governed by the
$k$th-order Hamiltonian is no longer adiabatic and
the effect of this nonadiabaticity may iteratively accumulate to affect the lower-order adiabaticity.
\widetext

\begin{table}[!hbp]
\begin{tabular}{|c|c|c|c|}
\hline
\textbf{Order} & \textbf{Deviations} & \textbf{Associated Hamiltonian} & \textbf{Adiabatic Parameters}  \\
\hline
0 & $\bar{p},\bar{q},|\bar{\psi}\rangle$ & $H_0(\hat{H}_0)$ & $R$ \\
\hline
1 & $\delta p,\delta q$ & $H_1$ & $R,\dot{R}$ \\
\hline
2 & $\delta^2 p,\delta^2 q$ & $H_2$ & $R,\dot{R},\ddot{R}$ \\
\hline
3 & $\delta^3 p,\delta^3 q$ & $H_3$ & $R,\dot{R},\ddot{R},d^3R/dt^3$ \\
\hline
\multicolumn{1}{|c|}{\textbf{\vdots}} & \multicolumn{1}{|c|}{\textbf{\vdots}} &\multicolumn{1}{|c|}{\textbf{\vdots}} &\multicolumn{1}{|c|}{\textbf{\vdots}} \\
\hline
$k$ & $\delta^k p,\delta^k q$ & $H_k$ & $R,\dot{R},\ddot{R},\ldots,d^kR/dt^k$ \\
\hline


\end{tabular}
\caption{Hierarchical adiabatic following at different orders, with the associated adiabatic Hamiltonians determined by various derivatives of the adiabatic parameter $R(t)$.}
\end{table}

\endwidetext

\begin{figure}[!htbp]
\vspace*{-0.5cm}
\centering
\includegraphics[width=0.45\textwidth]{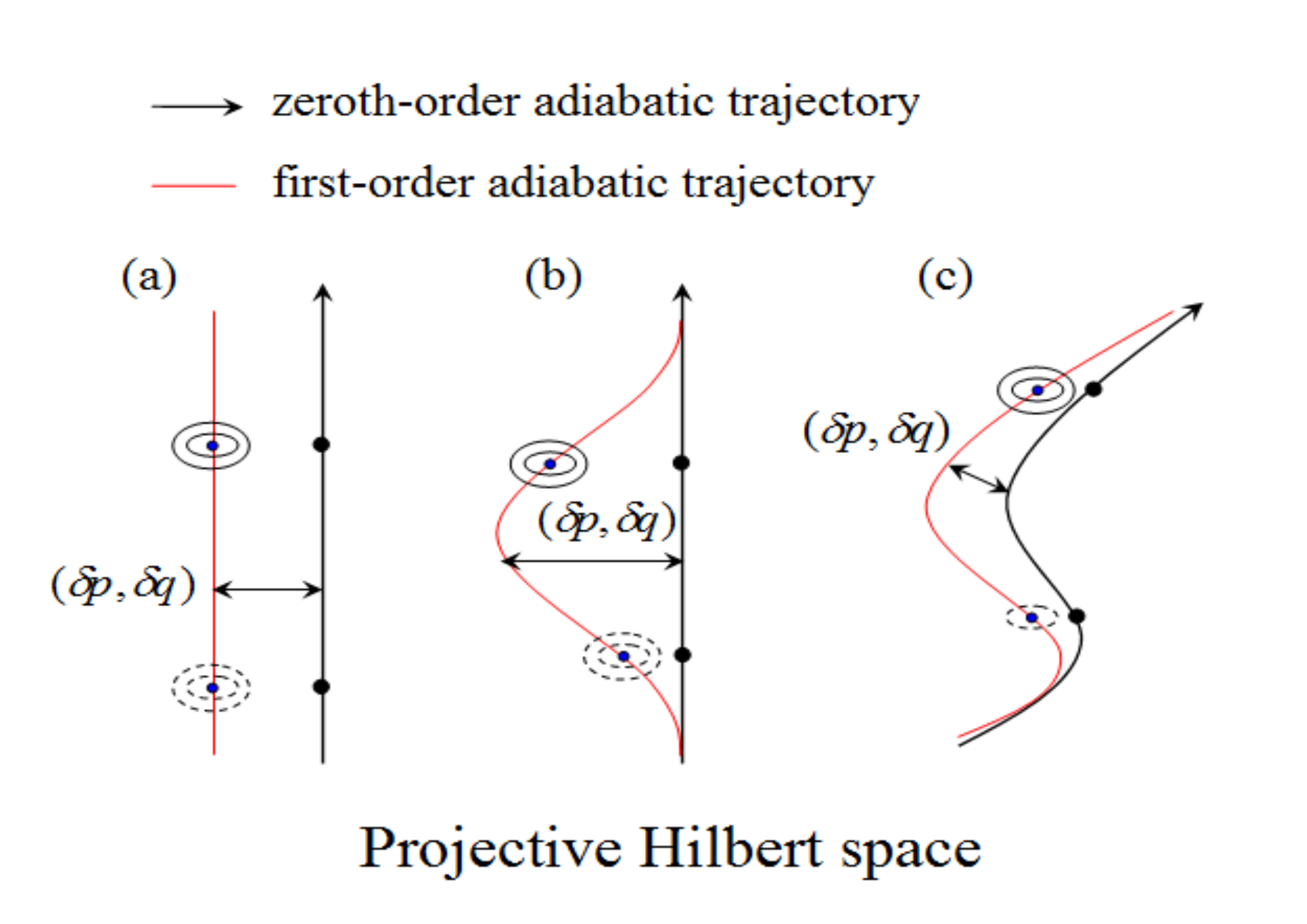}
\caption{(color online) Schematics of different scenarios for
the first-order deviation from the zeroth-order adiabatic trajectory.
(a) For the case of a spin-1/2 particle under a rotating field considered in Sec.~V.A,
 the first-order deviation from the zeroth-order adiabatic trajectory is a constant. (b) In the case of a LZ process considered in Sec.~V.B,  the average deviation in either the initial stage or the final stage approaches zero, but it can be appreciable at the intermediate stage. (c) The  first-order deviations $(\delta p, \delta q)$
can be manipulated by designing the time dependence of $R(t)$ and $\dot{R}(t)$. }
\label{first}
\end{figure}

In many practical systems for a limited time scale, it is sufficient to consider the first-order deviation, neglecting
all higher-orders. In Fig.~\ref{first}, we have depicted schematically three typical scenarios of the
first-order deviations. It is clear that the first-order deviations can be manipulated
by designing  $R(t)$ and $\dot{R}(t)$. This can be very useful to control
the nonadiabatic error in quantum adiabatic computation \cite{naQAC,naQAC1}.
We plan to pursue this issue in the near future. Moreover, by substituting the hierarchically corrected wavefunction into the original   Schr\"odinger equation, we may study possible corrections to the overall phase of the time-evolving quantum state \cite{liu,liu1}.

Our approach can be directly applied to classical adiabatic processes and nonlinear
quantum adiabatic evolution on the mean-field level \cite{ZhangPRL,nonlinear,nonlinear1,nonlinear2,ZhangAOP2012,ZhangJPA2012}.
For example, it is of considerable interest to apply our findings to assist in the control of adiabatic
processes in both classical and quantum systems \cite{Jar,Gong,Jar2}.

\vspace{0.3cm}
\section{Acknowledgments}
This work is supported by the NBRP of China (2013CB921903,2012CB921300) and
the NSF of China (11105123,11274024,11334001).

\bibliographystyle{apsrev}
\vskip10pt

\begin{appendix}

\section{Detailed derivations for the second-order theory}
The premise of dealing with the second-order deviation is that the state is around the first-order fixed point.
This allows us to  express $(\delta p, \delta q)$ as the following,
\begin{equation} \label{2nd-delta}
\delta p = \delta^2p +  A_1;  \ \ \ \delta q = \delta^2q +  B_1,
\end{equation}
where $\delta^2 p$ and $\delta^2 q$ describe the actual dynamics of $(\delta p, \delta q)$ on top of their time-averaged values $(A_1,B_1)$.

Note that in deriving $H_1$ we have only kept the first-order term when expanding the force field $\left[-\frac{\partial H_0 (R)}{\partial q}, \frac{\partial H_0(R)}{\partial p}\right]$. This is adequate for the first-order theory. When considering the second-order deviation, we should also keep the second-order terms in the expansion. Specifically,
substituting Eq.~(\ref{2nd-delta}) into Eqs.~(\ref{fd},\ref{afd}), keeping the second-order expansion terms
\begin{equation}
\frac{1}{2}\left(\delta p\frac{\partial}{\partial p}+\delta q\frac{\partial}{\partial q}\right)^2 H_0'  \quad\quad  H_0'=\frac{-\partial H_0}{\partial q} \quad {\text{or}} \quad \frac{\partial H_0}{\partial p},
\end{equation}
and neglecting terms containing $[\delta^2p]^2$ or $[\delta^2q]^2$(which are fourth-order), one finds (employing Eq.~(\ref{center}))
\begin{equation} \label{2nd-expansion}
\left(\begin{array}{c}\frac{d\delta^2p}{dt}\\ \\
\frac{d\delta^2q}{dt}\end{array}\right)=
\frac{1}{2}\delta\Gamma
 \left(\begin{array}{c} A_1\\ \\
 B_1\end{array}\right)+\left(\Gamma_0+\delta\Gamma\right)
 \left(\begin{array}{c} \delta^2p\\ \\
 \delta^2q\end{array}\right)
-\left(\begin{array}{c}\frac{dA_1}{dt}\\ \\
\frac{dB_1}{dt}\end{array}\right),
\end{equation}
where $\delta\Gamma$ is defined in Eq.~(\ref{shiftgamma}) as the state under consideration shifts from $(\bar{p},\bar{q})$ to $(\bar{p}+A_1,\bar{q}+B_1)$.

Rearranging some terms on the right-hand side of Eq.~(\ref{2nd-expansion}), we arrive at
\begin{eqnarray}
\left(\begin{array}{c}\frac{d\delta^2p}{dt}\\ \\
\frac{d\delta^2q}{dt}\end{array}\right)&=&\left(\Gamma_0+\delta\Gamma\right)\Bigg\{\left(\begin{array}{c} \delta^2p \\ \\ \delta^2q\end{array}\right)-\left(\Gamma_0+\delta\Gamma\right)^{-1}
\left(\begin{array}{c}\frac{d A_1}{d t}\\ \\
\frac{d B_1}{d t}\end{array}\right)   \nonumber \\
&&+\frac{1}{2}\left(\Gamma_0+\delta\Gamma\right)^{-1}\delta\Gamma
 \left(\begin{array}{c} A_1\\ \\
 B_1\end{array}\right)\Bigg\}.
\label{2nddynamics}
\end{eqnarray}
The fixed-point solution for $\delta^2p$ and $\delta^2q$ can be found from Eq.~(\ref{2nddynamics}); it is
\begin{eqnarray} \label{center2zapp}  \nonumber
\left(\begin{array}{c} A_2\\ \\
 B_2\end{array}\right) &=&\Gamma_0^{-1}(R)\left[\left(\begin{array}{c}\frac{\partial A_1}{\partial R}\\ \\
\frac{\partial B_1}{\partial R}\end{array}\right) \dot{R}+
\left(\begin{array}{c}\frac{\partial A_1}{\partial V}\\ \\
\frac{\partial B_1}{\partial V}\end{array}\right)
\ddot{R}\right] \\ &-&\frac{1}{2}\Gamma_0^{-1}\delta\Gamma\left(\begin{array}{c}A_1\\ \\
 B_1\end{array}\right),
\end{eqnarray}
where  the time derivatives of the adiabatic parameter $R$,  $\dot{R}$ and $\ddot{R}$,  are assumed to be in the same order of magnitude. All higher-order terms, such as those terms of the order of $\dot{R}^j$ with $j\geqslant3$, are neglected.  Under this treatment, it is now seen that, in terms of their time-averaged values, a
more accurate prediction of $(\delta p, \delta q)$ is given by $(A_1+A_2,B_1+B_2)$.  Note that $A_2$ and $B_2$ are evidently proportional to $\propto \dot{R}^2$. Equations (\ref{2nddynamics},\ref{center2zapp}) are just the second-order dynamics and the second-order fixed point given in the main text (see Eqs.~(\ref{2ndH}) and (\ref{center2z})).  One can now readily write down the second-order Hamiltonian
\begin{eqnarray} \label{Hflu2psu} \nonumber
H_2(R,\dot{R})&=&\frac{1}{2}\left(\frac{\partial^2H_0}{\partial q^2}\right)_{p_1,q_1}(\delta^2 q-B_2)^2 \\ \nonumber
&+&\left(\frac{\partial^2H_0}{\partial q\partial p}\right)_{p_1,q_1}(\delta^2 q-B_2) (\delta^2 p-A_2) \\
&+&\frac{1}{2}\left(\frac{\partial^2H_0}{\partial p^2}\right)_{p_1,q_1}(\delta^2 p-A_2)^2,
\end{eqnarray}
One only need to note that $(p,q)$ take value of $(\bar{p}+A_1,\bar{q}+B_1)$ instead of
$(\bar{p},\bar{q})$ as we are at the second-order approximation.

\section{High-order deviations in quantum adiabatic evolution}
The dynamics of the $k$th-order deviation $(\delta^kp,\delta^kq)$ can be derived iteratively
by substituting $\delta p=A_1+A_2+\ldots+\delta^kp$ and $\delta q=B_1+B_2+\ldots+\delta^kq$ into Eqs.~(\ref{fd},\ref{afd}) with the expansion up to the $k$th-order, provided the fixed points
of all the previous $(k-1)$ orders  have been obtained.
Specifically, $(\delta^3p,\delta^3q)$ can be described by a third-order Hamiltonian
$H_3(R, \dot{R}, \ddot{R},\frac{d^3R}{dt^3})$. The $k$th-order deviation $(\delta^kp,\delta^kq)$ forms a pair of canonical variables of a $k$th-order Hamiltonian $H_k(R, \dot{R}, \ddot{R},\ldots,\frac{d^kR}{dt^k})$, demonstrating that the $k$th-order deviation will undergo adiabatic evolution only if the time derivatives of parameter $R$ up to the $k$th-order are all manipulated very slowly in comparison with the intrinsic
frequency $\omega_k$ of the $k$th-order Hamiltonian, which is proportional to $|\Gamma_k|\approx|\Gamma_0|$.

The $k$th order deviation  consists of $k$ terms, with the first one associated with the ideal matrix $\Gamma_0$ and the adiabatic evolution of the $(k-1)$th-order deviation $(\delta^{k-1}p,\delta^{k-1}q)$, the second one associated with $\delta\Gamma$ and the evolution of $(\delta^{k-2}p,\delta^{k-2}q)$, and the $k$th one associated with $\delta^{k-1}\Gamma$ and the zeroth order adiabatic evolution of $(\bar{p}, \bar{q})$. The sum of the $k$ terms is the  result for the dynamical fixed point of $H_k$.

To illustrate that a general $k$th order theory is possible, we consider here only a rather simple case where $\dot{R}$ is a constant. However, even in this case our expressions appear to be complicated and hence readers may skip the technical details (we present them just for completeness). In particular, the $k$th-order
fixed point is   
\begin{equation} \label{centerk}
\left(\begin{array}{c} A_k\\ \\
 B_k\end{array}\right)=\Gamma_0^{-1}\left(\begin{array}{c}\frac{\partial A_{k-1}}{\partial R}\\ \\
 \frac{\partial B_{k-1}}{\partial R}\end{array}\right)\dot{R}
 -\Gamma_0^{-1}\sum_{j=1}^{k-1}\Delta^j\Gamma\left(\begin{array}{c} A_{k-j}\\ \\
 B_{k-j}\end{array}\right)\,.
\end{equation}
The deviations $\Delta^j\Gamma$ in Eq.~(\ref{centerk}) is defined as
\begin{equation} \label{long}
\Delta^j\Gamma={\cal T}^j\left\{\sum_{i=1}^j \frac{1}{(i+1)!} \left[\sum_{r=1}^j(A_r)\frac{\partial}{\partial p}+\sum_{r=1}^j(B_r)\frac{\partial}{\partial q}\right]^i\Gamma \right\}
\end{equation}
The function ${\cal T}^j(\ldots)$ in (\ref{long}) is to take the $j$th-order terms in $(\ldots)$, i.e., taking the sum of all the terms of the kind $A_t^uB_s^v$ with $tu+sv=j$. For example, $A_2$ and $B_2$ are second-order terms in terms of $\dot{R}$, and $A_2^2$ and $A_2B_2$ become the fourth-order terms, so ${\cal T}^2(A_2+B_2+A_2^2+A_2B_2)=A_2+B_2$, ${\cal T}^3(A_2+B_2+A_2^2+A_2B_2)=0$ and ${\cal T}^4(A_2+B_2+A_2^2+A_2B_2)=A_2^2+A_2B_2$, {\it etc}. Specifically, when $j=1$, $\Delta\Gamma=\frac{1}{2}\delta\Gamma$.

In the case of nonconstant adiabatic speed $V$, we should include the derivatives of the kind ($d^0R/dt^0\equiv R$)
\begin{equation} \nonumber
\sum_{j=0}^{k-1}\left(\begin{array}{c}\frac{\partial A_{k-1}}{\partial (d^jR/dt^j)}\\ \\
 \frac{\partial B_{k-1}}{\partial (d^jR/dt^j)}\end{array}\right) \cdot
 \frac{d^{j+1}R}{dt^{j+1}}
\end{equation}
for the $k$th-order deviation.

Generally, the hierarchy adiabatic theory can also be naturally extended to $n$-mode quantum system by expanding the $\Gamma$ matrix from dimension $2\times2$ to dimension $2(n-1)\times2(n-1)$.

Finally, it is necessary to make two remarks on  high-order deviations. First, the deviations of all orders are obtained  with respect to what the usual quantum adiabatic theorem predicts. This is the reason that Eq.~(\ref{long}) looks complicated. Second, in deriving the $k$th-order deviation in (\ref{centerk}), we have already assumed the adiabaticity holds for up to the $k$th-order Hamiltonian.

\end{appendix}

\end{document}